# Folksodriven Structure Network

Massimiliano Dal Mas

*me @ maxdalmas.com*

**Abstract.** Nowadays folksonomy is used as a system derived from user-generated electronic tags or keywords that annotate and describe online content. But it is not a classification system as an ontology. To consider it as a classification system it would be necessary to share a representation of contexts by all the users. This paper is proposing the use of folksonomies and network theory to devise a new concept: a "Folksodriven Structure Network" to represent folksonomies. This paper proposed and analyzed the network structure of Folksodriven tags thought as folsksonomy tags suggestions for the user on a dataset built on chosen websites. It is observed that the Folksodriven Network has relative low path lengths checking it with classic networking measures (clustering coefficient). Experiment result shows it can facilitate serendipitous discovery of content among users. Neat examples and clear formulas can show how a "Folksodriven Structure Network" can be used to tackle ontology mapping challenges.

## 1 Introduction

The communicative form of the World Wide Web is based on user-centric publishing and knowledge management platforms as sharing systems for social and collaborative matching like: Wikis, Blogs, Facebook, etc... Ontology defines a common set of sharing concepts, but unfortunately ontologies are not wide spread at the moment. While Folksonomy is said to provide a democratic tagging system that reflects the opinions of the general public, but it is not a classification system and it is difficult to make sense of [1]. The emergent labeling of lots of resources by untrained people can be used by different kind of tools that can benefit for the individual user. A representation of contexts should be share by all the users. The goal of this work is to help the users to choose proper tags thanks to a "Folksodriven Structure Network" intended as a dynamical driven system of folksonomy that could evolve during the time. In this work the main network characteristics was analyzed by a group of articles from a chosen websites and analyzed according to the Natural Language Processing. The data structures extracted is represented on folksonomy tags that are correlated with the source and the relative time exposition - measure of the time of its disposal. Considering those we define a tag structure called Folksodriven, and adapt classical network measures to them.

## 2 Folksodriven Notation

$$(1) \quad FD := (C, E, R, X)$$

A Folksodriven will be considered as a tuple (1) defined by finite sets composed by:

- *Formal Context* (C) is a triple $C := (T, D, I)$ where the objects $T$ and the attributes $D$ are sets of data and $I$ is a relation between $T$ and $D$ [2] – see 3.2 (*Incidence of context on folksonomy technology*);
- *Time Exposition (E)* is the clickthrough rate (CTR) as the number of clicks on a *Resource (R)* divided by the number of times that the *Resource (R)* is displayed (impressions);
- *Resource (R)* is represented by the uri of the webpage that the user wants to correlate to a chosen tag.
- *X* is defined by the ternary relation $X = C \times E \times R$ in a Minkowski vector space delimited by the vectors *C*, *E* and *R*. [1]

---

An article A can be represented as a relation according to the Folksodriven sets:

$$(2) \quad A(c,r) := \{(c,e,r) \in X \mid e \in E\}$$

## 3 Folksodriven Data Set

The data set has been built from articles taken from web sites news for a period of one month, because they are frequently updated. Tokens will be extracted from the title (*T*) and description (*D*) of the articles. Those tokens compose a data set of words proposed to the users as Tags that he/she can add to a document - the articles on the web sites - to describe it. Chunking was used in this work as a starting point but it is at a very low semantic level. [2]

### 3.2 Incidence of context on folksonomy technology

In this paper the notion *context* is used in the sense of *formal context* as used in the ontological sense defined by the Formal Concept Analysis (FCA) - a branch of Applied Mathematics [5] - for the dynamic corpus on chunking operation.

$$(3) \quad C_n := (T_n, D_n, I_n)$$

A *set of formal contexts C* is defined by (3) considering:
- *T* as a *set of title tags;*
- *D* as a *set of description tags D*;
- *I* as a *set of incidence relations of context,* defined by the frequency of occurrence of the relation between *T* and *D* as depicted in (4),

The tag *T* derived by the title was considered as a facet described by the tag *D* derived by the description. On (3) the *set of incidence relations of context I* is defined by the matching between *T* and *D* tags by relation (4) allowing multiple associations among *D* tags and the faceted context defined by every *T* tag.

$$(4) \quad I \subseteq T \times D$$

Multiple matching was disambiguated by updating a Jaccard similarity coefficient associated with the incidence relation of context. [3]

In this way a selected number of chunks, defined according to the *Formal Context* (*C*), are proposed to the user as folksonomy tags for the correlated uri *Resource* (*R*). So the "Folksodriven Data Set" can "drive" the user on the choice of a correct folksonomy tag.

---

1. In a model of space-time, every point in space has four coordinates (*x, y, z, t*), three of which represent a point in space, and the fourth a precise moment in time. Intuitively, each point represents an event that happened at a particular place at a precise moment. The usage of the *four-vector* name assumes that its components refer to a "standard basis" on a Minkowski space [3]. Points in a Minkowski space are regarded as events in space-time.

2. Chunking (or Shallow Parsing) is the process of identifying syntactical phrases in natural language sentences. Identifying whole parse trees can provide deeper analyses of the sentences but it can be a hard problem respect identifying some segments of a text. A chunking operation segments text into an unstructured sequence of text units called "chunks" [4]. All chunks are represented at the same flat level non-overlapping regions of text (discontinuous chunks are not allowed) and been non-recursive (chunks containing another chunk of the same category are not allowed) [4].

3. Jaccard similarity coefficient [6], also known as the Tanimoto coefficient, measures the overlap of two sets. It is zero if two sets of the *incidence relation of context* are disjoint (i.e., they have no common members) and is one if they are identical: $J(T,D) = |T \cap D| / |T \cup D|$

## 4 Folksodriven as a Network

The Folksodriven tags can be depicted by network patterns in which nodes are Folksodriven tags and links are semantic acquaintance relationships between them according to the SUMO (http://www.ontologyportal.org) formal ontology that has been mapped to the WordNet lexicon (http://wordnet.princeton.edu).

It is easy to see that Folksodriven tags tend to form groups (as small groups in which tags are close related to each one) and Folksodriven tags of a group also have a few acquaintance relationships to Folksodriven tags outside that group. Folksodriven tags can be related to other tags (e.g., workers, engineers) that are connected to other groups. In this way Folksodriven tags may be considered the hubs responsible for making such network a "scale-free network". In a scale-free network most nodes of a graph are not neighbors of one another but can be reached from every other by a small number of hops or steps, considering the mutual acquaintance of Folksodriven tags. [7]

### 4.1 Folksodriven Clustering Coefficient

An important characteristic of scale-free networks is the *Clustering coefficient* distribution, which decreases as the node degree increases following a power law. [7]

*Clustering coefficient* is a local property measuring the connectedness of the neighbours (as defined by Watts [8]). We consider an exclusive vs. an overlapping clustering [4] as the ratio between the maximum and the minimum value connectedness of the neighbours of a Folksodriven tag to the uri resource $r$ considered.

$$(5) \quad K_r(i) := \frac{|C_r(i) \times E_r(i)|}{|C_r(i)| \bullet |E_r(i)|}$$

According to the (5) it is possible to see how a high value of the *Clustering coefficient $K_r(i)$* indicates that a great number of users are increasing the $E_r(i)$ to a $C_r(i)$ for the same uri resource $r$.

## 5 Experiments

### 5.1 Setup

A test network model was realized in a simulated environment to check the Scale-free Network structure of the Folksodriven tags.

The Scale-free Network was compared with a random graph generated adding tags one at a time joining to a fixed number of starting tags, that are chosen with probability proportional to the graph degree - this model of the growth of the World Wide Web was developed by Barabasi and Albert [7].

All data were obtained from averages over 100 identical network realizations with a sample of 400 nodes taken randomly from each graph performing twenty runs to ensure consistency.

### 5.2 Data set analysis

Figures 1-2 show the results for the Clustering for the dataset. We consider the *Clustering coefficients* ($K_r$) as a function of the network degree.

The characteristic *Folksodriven Clustering Coefficient* (Figure 1) is considerably smaller than for the random graph. The *Clustering Coefficient* has remained almost constant at about 2.5 while the number of nodes has grown about twenty during the observation period. This imply that on average, every *Formal Context (C)*, *Time Exposition (T)* and *Resource (R)* defined on the original data set can be reached within 2.5 mouse clicks from any given page.

We can argue that everything in it is still reachable within few hyperlinks even if the Folksodriven graph grows to millions of nodes. This attest the context of "serendipitous discovery" [9] of contents in the folksonomy community.

---

[4] Exclusiveness strategy: each object can be assigned to only one cluster at a time allowing a minimum connectedness of the neighbours of the Folksodriven tag referred to the uri resource $r$.

Overlapping strategy: there are multiple assignments of any object allowing a maximum connectedness of the neighbours of a Folksodriven tag referred to the uri resource $r$.

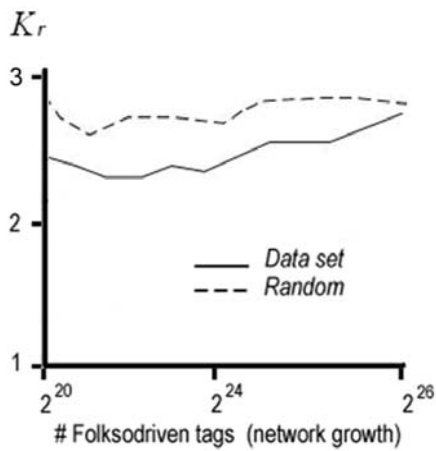 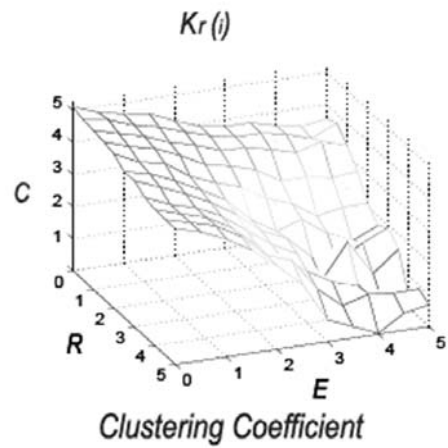

**Figure 1:** Set of data compared with the corresponding Random graphs for Folksodriven *Clustering coefficient*. It is depicted how the characteristic path length takes quite similar values for the corresponding Random graph.

**Figure 2:** *Clustering coefficient* is depicted in the space delimited by $C$, $E$, and $R$.. For larger time exposition $E$ the *Clustering coefficients* become drastically smaller, as expected for the $E \to \infty$ and $C \to 0$ limit.

# 6 Future work

Web is full of information that often lacks a systematic reference. The Folksodriven structure can determine a user contribution [10], so in a community environment will be possible to make explicit groups of interests to be used for: recommender systems, opinion mining, and sentiment analysis.
Possible applications for such framework based on that new method can be employed by different kinds of knowledge management systems that need to use the knowledge shared by different people like.

**Massimiliano Dal Mas** is an engineer at the Web Services division of the Telecom Italia Group, Italy. His interests include: user interfaces and visualization for information retrieval, automated Web interface evaluation and text analysis, empirical computational linguistics, and text data mining. He received BA, MS degrees in Computer Science Engineering from the Politecnico di Milano, Italy. He won the thirteenth edition 2008 of the CEI Award for the best degree thesis with a dissertation on "Semantic technologies for industrial purposes" (Supervisor Prof. M. Colombetti).